# The basic Leggett inequalities don't contradict the quantum theory, neither the classical physics

*Sofia Wechsler* [1)]

**Abstract**

The basic Leggett inequalities, i.e. those inequalities in which the particular assumptions of Leggett's hidden-variable model (e.g. Malus law) were not yet introduced, are usually derived using hidden-variable distributions of probabilities (although in some cases completely general, positive probabilities would lead to the same result). This fact creates sometimes the illusion that these basic inequalities are a belonging of the hidden-variable theories and are bound to contradict the quantum theory.
In the present text the basic Leggett inequalities are derived in the most general way, i.e. no assumption is made that the distribution of probabilities would result from some wave function, or from some set of classical variables. The consequence is that as long as one and the same probability distribution is used in the calculus of all the averages appearing in the basic Leggett inequalities, no contradiction may occur. These inequalities may be violated only when different averages are calculated with different distributions, for example, some of them calculated with the quantum formalism and the others with some distribution of classical parameters.

## 1. Introduction

Since Leggett's hidden-variable (HV) model for entanglements was published [1], a series of experiments were performed to prove a conflict between this model and the quantum theory [2, 3]. The development of the model begins with two very general inequalities, [4], containing averages on two observables. They are referred to below as the *basic Leggett inequalities*, to distinguish them from the final inequalities obtained by Leggett, which conflict with the quantum theory. The conflict arises because some of the averages appearing in the basic inequalities were further calculated using the quantum formalism, while the other averages were calculated using the distribution of hidden parameters in Leggett's model.

There is a misleading fact about the basic Leggett inequalities: they are usually derived in the literature applying to a HV distribution, see for instance [5], while completely general and positive probabilities would obviously lead to the same result. This fact creates sometimes the illusion that these inequalities are a belonging of the HV theories, and are therefore bound to contradict the quantum theory.
In this text a derivation of the basic Leggett's inequalities is shown, using a completely general distribution of positive probabilities of the observable results. The derivation below is very detailed to give the reader the possibility to check at any step that no additional assumptions are introduced. Besides this particularity this derivation is not a novelty. Leggett's own derivation of these inequalities was very general, [4].

## 2. A general proof

Let's consider as in [1] two observables, *A* and *B*, that take only the values ±1. They satisfy the equality,

(1) $1 - |A - B| = AB = -1 + |A + B|$,

that may be easily checked for each one of the combinations of values of *A* and *B*, i.e. $A = B = +1$, $A = -B =$

---
[1)] Computers Engineering Center, Nahariya, P.O.B. 2004, 22265, Israel

+1, etc. Let $P_{AB}(A, B; a, b)$ be the joint probability distribution of the values of $A$ and $B$ for some particular settings, $a$ and $b$, of the apparatuses; for example, in experiments with pairs of polarized particles, $a$ ($b$) may be the direction on which is tested the polarization of the particle 1 (2) in a given trial of the experiment. A short notation $P_{AB}(+, +)$ for $P_{AB}(A = +1, B = +1; a, b)$, $P_{AB}(+, -)$ for $P_{AB}(A = +1, B = -1, a, b)$, etc., will be used in the rest of the article.

Now, multiplying all the sides of (1) with each one of the four joint probabilities, one obtains

$$P_{AB}(+, +) = P_{AB}(+, +) = -P_{AB}(+, +) + 2P_{AB}(+, +),$$

$$P_{AB}(-, +) - 2P_{AB}(-, +) = -P_{AB}(-, +) = -P_{AB}(-, +),$$

$$P_{AB}(+, -) - 2P_{AB}(+, -) = -P_{AB}(+, -) = -P_{AB}(+, -),$$

$$P_{AB}(-, -) = P_{AB}(-, -) = -P_{AB}(-, -) + 2P_{AB}(-, -).$$

Adding these four equalities there results

$$(2)\quad 1 - 2P_{AB}(+, -) - 2P_{AB}(-, +) = \overline{AB} = -1 + 2P_{AB}(+, +) + 2P_{AB}(-, -).$$

Since probabilities are positive numbers as said above, the following inequalities hold for the leftmost side

$$(3)\quad 1 - 2P_{AB}(+, -) + 2P_{AB}(-, +) \geq 1 - 2P_{AB}(+, -) - 2P_{AB}(-, +) \quad (a),$$

$$1 + 2P_{AB}(+, -) - 2P_{AB}(-, +) \geq 1 - 2P_{AB}(+, -) - 2P_{AB}(-, +) \quad (b).$$

Comparing (3a) and (3b) with (2) one gets respectively

$$(4)\quad 1 - 2P_{AB}(+, -) + 2P_{AB}(-, +) \geq \overline{AB} \quad (a),$$

$$1 - 2P_{AB}(+, -) + 2P_{AB}(-, +) \geq \overline{AB} \quad (b).$$

By adding and subtracting $P_{AB}(+, +)$ and $P_{AB}(-, -)$ and re-arranging terms, the LHS of (4a) becomes:

$$1 - [P_{AB}(+, +) + P_{AB}(+, -) - P_{AB}(-, +) - P_{AB}(-, -)] + [P_{AB}(+, +) - P_{AB}(+, -) + P_{AB}(-, +) - P_{AB}(-, -)]$$

$$= 1 - \overline{A} + \overline{B}.$$

After a similar treatment, the LHS of (4b) becomes $1 + \overline{A} - \overline{B}$. Therefore (4) takes the form

$$(5)\quad 1 - \overline{A} + \overline{B} \geq \overline{AB} \quad (a), \qquad 1 + \overline{A} - \overline{B} \geq \overline{AB} \quad (b).$$

These two inequalities prove that

$$(6)\quad 1 - |\overline{A} - \overline{B}| \geq \overline{AB}.$$

Indeed, if $\overline{A} - \overline{B} > 0$, it's easy to see that the LHS of (6) is equal to the LHS of (5a), while if $\overline{A} - \overline{B} < 0$, the LHS of (6) is equal to the LHS of (5b).

Now we return to the equality (2) and examine its rightmost side. Since probabilities are positive, the following inequalities hold

(7) $-1 + 2P_{AB}(+, +) + 2P_{AB}(-, -) \geq -1 + 2P_{AB}(+, +) - 2P_{AB}(-, -)$ (a),

$-1 + 2P_{AB}(+, +) + 2P_{AB}(-, -) \geq -1 - 2P_{AB}(+, +) + 2P_{AB}(-, -)$ (b).

Comparing (7a) and (7b) with (2) one gets respectively

(8) $\overline{AB} \geq -1 + 2P_{AB}(+, +) - 2P_{AB}(-, -)$ (a), $\overline{AB} \geq -1 - 2P_{AB}(+, +) + 2P_{AB}(-, -)$ (b).

By adding and subtracting $P_{AB}(+, -)$ and $P_{AB}(-, +)$ and re-arranging terms, the RHS of (8a) becomes:

$$-1 + \left[P_{AB}(+, +) + P_{AB}(+, -) - P_{AB}(-, +) - P_{AB}(-, -)\right] + \left[P_{AB}(+, +) + P_{AB}(-, +) - P_{AB}(+, -) - P_{AB}(-, -)\right]$$

$$= -1 + \bar{A} + \bar{B}.$$

After a similar treatment, the RHS of (8b) becomes $-1 - \bar{A} - \bar{B}$. Therefore (8) takes the form

(9) $\overline{AB} \geq -1 + \bar{A} + \bar{B}$ (a), $\overline{AB} \geq -1 - \bar{A} - \bar{B}$ (b).

These two inequalities prove that

(10) $\overline{AB} \geq -1 + |\bar{A} + \bar{B}|$.

Indeed, if $\bar{A} + \bar{B} > 0$, the RHS of (10) is equal to the RHS of (9a), while if $\bar{A} + \bar{B} < 0$, the RHS of (10) is equal to the RHS of (9b).

Finally, (10) and (6) may be combined into

(11) $1 - |\bar{A} - \bar{B}| \geq \overline{AB} \geq -1 + |\bar{A} + \bar{B}|$,

which are the basic Leggett inequalities.

**Note:** while completing this article I became aware that R. Lapiedra was about to provide a similar proof, in correction of an erroneous rationale in his quant-ph/0806.0307 after the criticism and advice of different readers.

## Acknowledgement

I am deeply thankful to Dr. M. P. Seevinck for calling my attention on the fact that Leggett's own proof of the above discussed inequalities was also completely general, and on the fact that the first version of this article wrongly used the term "Leggett inequalities" for Leggett's basic inequalities.